\documentclass[twocolumn]{aastex631}

\usepackage{color}
\usepackage{amsmath}
\shorttitle{Multi-Zone Blazar Polarization}
\shortauthors{Zhang et al.}

\begin{document}

\title{Revisiting High-Energy Polarization from Leptonic and Hadronic Blazar Scenarios}

\correspondingauthor{Haocheng Zhang}
\email{haocheng.zhang@nasa.gov}

\author[0000-0001-9826-1759]{Haocheng Zhang}
\affiliation{University of Maryland Baltimore County\\
Baltimore, MD 21250, USA}
\affiliation{NASA Goddard Space Flight Center\\
Greenbelt, MD 20771, USA}

\author[0000-0002-8434-5692]{Markus B\"ottcher}
\affiliation{Centre for Space Research\\
North-West University\\
Potchefstroom, 2531, South Africa}

\author[0000-0001-9200-4006]{Ioannis Liodakis}
\affiliation{NASA Marshall Space Flight Center\\
Huntsville, AL 35808, USA}



\begin{abstract}
X-ray and MeV polarization can be powerful diagnostics for leptonic and hadronic blazar models. Previous predictions are mostly based on a one-zone framework. However, recent IXPE observations of Mrk~421 and 501 strongly favor a multi-zone framework. Thus, the leptonic and hadronic polarization predictions need to be revisited. Here we identify two generic radiation transfer effects, namely, double depolarization and energy stratification, that can have an impact on the leptonic and hadronic polarization. We show how they are generalized from previously known multi-zone effects of the primary electron synchrotron radiation. Under our generic multi-zone model, the leptonic polarization degree is expected to be much lower than the one-zone prediction, unlikely detectable in most cases. The hadronic polarization degree can reach a value as high as the primary electron synchrotron polarization during simultaneous multi-wavelength flares, consistent with the one-zone prediction. Therefore, IXPE and future X-ray and MeV polarimeters such as eXTP, COSI, and AMEGO-X, have good chances to detect hadronic polarization during flares. However, the hadronic polarization cannot be well constrained during the quiescent state. Nonetheless, if some blazar jets possess relatively stable large-scale magnetic structures, as suggested by radio observations, a non-trivial polarization degree may show up for the hadronic model after a very long exposure time ($\gtrsim 1$ year).
\end{abstract}



\section{Introduction} \label{sec:intro}

Blazars are relativistic plasma jets from supermassive black holes that point very close to our line of sight. Due to relativistic beaming effects, they appear very bright in all electromagnetic wavelength bands, and are the most numerous extragalactic $\gamma$-ray sources seen by Fermi-LAT \citep{Fermi2020}. Their emission is highly variable, with the variability time scale in $\gamma$-rays reaching down to a few minutes \citep{Albert2007,Ackermann2016}. It is generally believed that the highly variable emission originates from an unresolved region, referred to as the blazar zone, which is located somewhere between sub-parsec and a few parsecs from the central engine \citep[for a recent review of the blazar physics, see][]{Boettcher2019}. Blazars exhibit characteristic, non-thermal-dominated two-component spectral energy distributions (SEDs). The low-energy spectral component, typically extending from radio to the optical bands, in some cases up to X-rays, is dominated by the synchrotron emission from nonthermal electrons. Thus, it is often referred to as the synchrotron component. This is evident by the observed high polarization in radio and optical bands \citep[e.g.,][]{Scarpa1997,Smith2007}, consistent with electron synchrotron emission in a partially ordered magnetic field \citep[e.g.,][for a recent review, see \citealp{ZHC2019b}]{Lyutikov2005,ZHC2015}. Blazars can be categorized into three types based on the peak frequency of the synchrotron component: low-synchrotron-peaked (LSP, which peaks below the optical band), intermediate-synchrotron-peaked (ISP, which peak around the optical and ultraviolet bands), and high-synchrotron-peaked (HSP, which peak in the X-ray band).

The high-energy spectral component extends from X-rays to the $\gamma$-ray band. Its origin is less understood. Leptonic models suggest that the high-energy emission comes from the inverse Compton scattering of low-energy photons (called the seed photons) by the same nonthermal electrons that produce the synchrotron component \citep{Marscher1985,Maraschi1992,Dermer1992,Sikora1994}. The seed photons can be the synchrotron photons themselves (referred to as synchrotron self Compton, SSC) or external photons from other sources (referred to as the external Compton, EC), such as thermal photons reprocessed in the broad line region, molecular clouds, and the dusty torus. Hadronic models, on the other hand, suggest that nonthermal protons can contribute to X-ray and $\gamma$-ray emission via proton synchrotron and photomeson processes, potentially triggering electromagnetic cascades \citep{Mannheim1992,Mucke2001}. Under this scenario, the high-energy emission does not necessarily correlate with the synchrotron component as the two components originate from different particle populations. One-zone spectral models suggest that the two scenarios can produce very similar SED fitting, although the hadronic models usually require a much higher energy budget \citep{Boettcher2013,Cerruti2015,Zdziarski2015,Liodakis2020}. Multi-wavelength observations have shown that the synchrotron and high-energy components are often correlated in many flaring events, generally supporting the leptonic scenario \citep[e.g.,][]{Rani2013,Liodakis2018,Liodakis2019}. However, the recent IceCube detection of a very high energy neutrino event in coincidence with the flaring blazar TXS~0506+056 and mounting evidence for a systematic spatial correlation between IceCube neutrino events and $\gamma$-ray loud blazars hint at a hadronic origin \citep{IceCube2018}. But existing models encounter difficulties in simultaneously explaining the multi-wavelength SEDs and neutrino emission from TXS~0506+056 and other blazars \citep[e.g.,][]{Cerruti2019,Keivani2018,Petropoulou2020}. The true origin of the high-energy component therefore remains unclear.

High-energy polarization can distinguish the leptonic and hadronic models \citep{ZHC2013}. This is because the synchrotron emission from either protons or charged secondary particles from hadronic cascades can be as highly polarized as the electron synchrotron emission that constitutes the low-energy component. If, on the other hand, X-ray and $\gamma$-ray emission is leptonic, isotropic inverse Compton scattering, which is generally expected in the comoving frame of the blazar zone, can reduce the polarization to about 40\% to 50\% of the seed photons \citep{Bonometto1973,Krawczynski2012}. In a perfectly ordered magnetic field, typical blazar parameters yield a maximal SSC polarization degree of $\sim 40\%$. EC is unpolarized since the seed thermal photons are usually unpolarized. However, radio and optical polarimetry reveals that the magnetic field is only partially ordered in the blazar zone \citep[e.g.,][]{Marscher2010,Blinov2021,Itoh2016}.The predicted X-ray and $\gamma$-ray polarization degrees thus need to be corrected, as suggested and used in a number of papers \citep{Paliya2018,ZHC2019,Schutte2022}, by applying an ordering factor, which is the ratio of the observed synchrotron polarization degree over the theoretical maximal polarization degree. For instance, if the observed optical polarization degree is 10\%, and the theoretical maximum is 70\%, then the ordering factor is $1/7$, and the predicted SSC and proton/pair synchrotron polarization degrees are $\sim 5\%$ and $\sim 10\%$, respectively.

However, the above estimate only applies to one-zone models. Recent IXPE observations of HSPs have shown that the optical and X-ray polarization degrees can be drastically different from each other \citep{Liodakis2022,DiGesu2022}. Even polarization angle swings, which often indicate major variations in the magnetic field, do not happen simultaneously in the optical and X-ray bands \citep{DiGesu2023,Middei2023}. These observations suggest that nonthermal particles of different energy are not strictly co-spatial. Instead, they are energy-stratified, which strongly argues for a multi-zone description. As a result, the polarization of the blazar high-energy component needs to be revisited under the multi-zone picture. Here we present two generic radiation transfer effects, double depolarization and energy stratification, in the multi-zone framework. They can significantly alter previous one-zone predictions for high-energy polarization. Section \ref{sec:ssc} describes the double depolarization effect and Section \ref{sec:hadronic} presents the energy stratification effect. Section \ref{sec:obs} discusses new predictions for IXPE, eXTP, and future MeV polarimeters such as COSI and AMEGO-X.

\section{Double Depolarization of SSC \label{sec:ssc}}

This section describes the double depolarization effect that can affect the SSC in an emission region with inhomogeneous magnetic fields. We suggest that if the magnetic field is completely random, the synchrotron polarization degree is proportional to $1/\sqrt{N}$, but the SSC should be scaled as $1/N^{5/6}$, where $N$ is the number of patches with uncorrelated magnetic field in the emission region. The polarization of synchrotron and SSC can be equally variable, but their temporal evolution can be different.

\subsection{Review of the Ordering Factor of Synchrotron Polarization}

We first review the synchrotron polarization degree in a blazar zone with inhomogeneous magnetic fields. Assuming that the blazar zone consists of $N$ incoherent patches of equal luminosity $L$ and linear polarization degree $\Pi_0$ but random polarization angles, since our distance to the blazar zone $d$ is much larger than the size of the blazar zone, each patch will contribute the same flux $F=L/(4\pi d^2)$ to the total flux and a segment of the same length $\Pi_0 F$ in the Stokes $Q$-$U$ plane. Since the polarized flux from different patches is incoherent, their Stokes parameters can be directly added up. The sum becomes a classical mathematical problem: a random walk in the $Q$-$U$ plane with the same step $\Pi_0 F$ for $N$ steps. The root mean square distance, which is also the average $\sqrt{Q^2+U^2}$, is simply $\sqrt{N}\Pi_0 F$. Thus the total polarization degree of the blazar zone is given by
\begin{equation}
\Pi_{tot}=\frac{\sqrt{Q^2+U^2}}{I}=\frac{\sqrt{N}\Pi_0 F}{NF}=\frac{1}{\sqrt{N}}\Pi_0 ~~.
\label{eq:pol}
\end{equation}
Clearly, if the emission from each patch is due to synchrotron by the same power-law distribution of electrons in a perfectly ordered magnetic field, then $\Pi_0=\Pi_{syn,max}$, where $\Pi_{syn,max}$ is the theoretical maximum of the synchrotron polarization degree for the given electron distribution. Hence,
\begin{equation}
\Pi_{syn,tot}=\frac{1}{\sqrt{N}}\Pi_{syn,max} ~~.
\label{eq:syn}
\end{equation}
The $1/\sqrt{N}$ is the theoretical basis for the aforementioned ordering factor of the magnetic field. Here we call this factor the primary depolarization factor.

The above is only an approximation mainly due to four issues. Firstly, random magnetic fields do not correspond to random polarization angles. This is because the synchrotron flux and polarization depend on the projected magnetic field onto the plane of sky instead of the actual magnetic field. Therefore, if the blazar zone has the same electron distribution and magnetic field strengths everywhere but the magnetic field directions are random, the magnetic field component that is parallel to the line of sight in the comoving frame can be considered as not contributing to the synchrotron flux and polarization. Such a blazar zone then does not match the equal luminosity and polarization degree, but random polarization angle assumptions used in deriving Equations \ref{eq:pol} and \ref{eq:syn}. Secondly, completely random magnetic fields are generally unrealistic. Take turbulence as an example, particle-in-cell (PIC) simulations of magnetized turbulence have found that the synchrotron polarization degree can be as high as 5\% to 10\%, even if the spatial resolution of the simulation domain in each dimension is larger than 1000 \citep{ZHC2023}. Similarly, the turbulent extreme multi-zone model (TEMZ), based on a Kolmogorov spectrum, finds the same polarization degree range \citep{Marscher2014}. This is because the magnetic field structure and evolution are not completely uncorrelated. Rather, turbulence can be described by the correlation length $\lambda_B$, so that magnetic fields on scales larger than the correlation length can be considered as uncorrelated. Thus, $N$ is the third power of the ratio of the size of the blazar zone $R$ over the correlation length of the turbulence, $N=(R/\lambda_B)^3$. Consequently, $N$ should be interpreted as the number of patches whose magnetic fields are uncorrelated in the blazar zone instead of the total simulation cells $N_{tot}$. In this case, the equal luminosity and polarization degree assumptions generally do not hold, but they depend on the physical conditions and particle acceleration mechanisms in the blazar zone. Thirdly, both observations and theories have shown that the magnetic field and particle distributions can change rapidly in the blazar zone \citep[e.g.,][]{Marscher2010,Blinov2021,ZHC2018}. Light crossing time effects can significantly alter the observed polarization signatures as shown in previous works \citep{Marscher2014,ZHC2015,ZHC2016b}. Thus, Equations \ref{eq:pol} and \ref{eq:syn} are only steady-state approximations. Finally, the root mean square used in deriving Equation \ref{eq:pol} is a statistical value, generally requiring $N\gg 1$ and $N_{tot}\gg N$. If the first condition is not met, the randomness of magnetic fields becomes dominating and the total polarization degree can be an arbitrary value below $\Pi_{syn,max}$. For the second condition, if the magnetic field in each patch is perfectly ordered, i.e., $\Pi_0=\Pi_{syn,max}$, then Equation \ref{eq:syn} holds as long as $N_{tot}\geq N$; but if the magnetic field has some structures in  each patch, so that $\Pi_0<\Pi_{syn,max}$, then without satisfying $N_{tot}\gg N$, the magnetic field structure in each patch is not well resolved, resulting in overestimate of the total polarization degree.

\subsection{Model for Double Depolarization of SSC}

\begin{figure}
\centering
\includegraphics[width=0.99\linewidth]{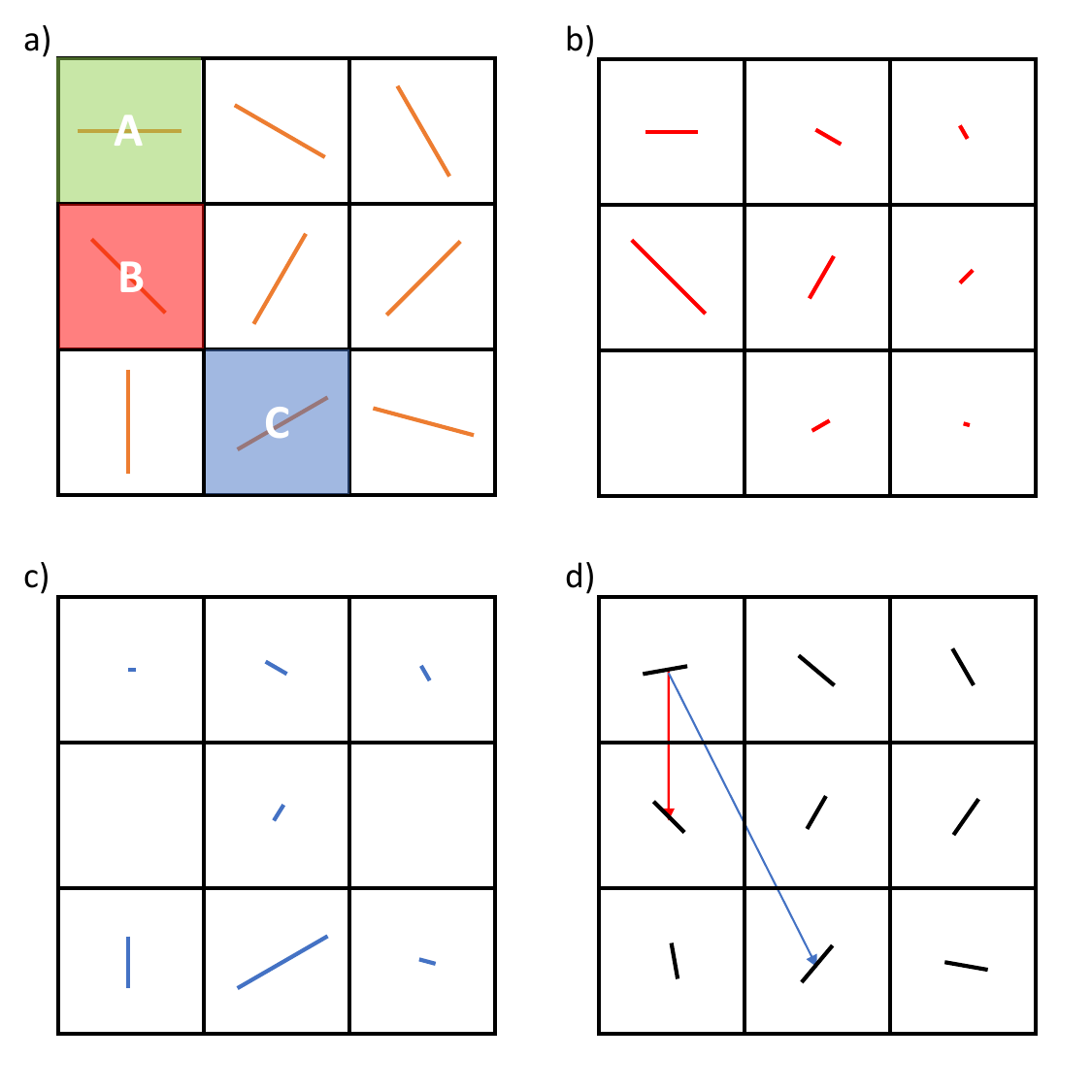}
\caption{A sketch of the double depolarization effect for SSC in the multi-zone picture. Panel a) shows the synchrotron polarized flux from nine patches. The magnetic field is in the paper plane and the viewing angle is perpendicular to the paper plane. The directions of the orange segments represent the polarization angle, which is perpendicular to the magnetic field direction, and their size represents the polarized flux amplitude. Three Patches A, B, C are highlighted in red, blue and green. Panel b) and c) show the SSC polarized flux contributed by each patch if the SSC happens in Patches B and C, respectively. Panel d) shows the total SSC polarized flux from all patches. The red and blue arrows from Patch A to Patches B and C, respectively, represent the internal light crossing time delays.}
\label{fig:sscsketch}
\end{figure}

\citet{Peirson2019} presents a multi-zone study of the SSC polarization with a conical jet model. Very interestingly, the paper finds that the SSC polarization degree is about 30\% of that of the synchrotron, which is lower than the usual $\sim 50\%$ found in previous works \citep{Bonometto1973,Krawczynski2012,ZHC2013}. The paper attributes this difference to the multi-zone effects under their conical jet model. Inspired by this, here we propose a generic radiation transfer effect that SSC can suffer from double depolarization in the multi-zone framework. This effect should apply if the blazar zone is optically thin to both the synchrotron seed photons and Compton scattered photons, which is typical for most part of the blazar spectra. Figure \ref{fig:sscsketch} gives a simple illustration of this effect. Suppose that the emission region has $N$ patches of uncorrelated magnetic field ($N=9$ in the sketch) that is well resolved ($N_{tot}\gg N$). We also assume that the magnetic field is fully in the paper plane and the viewing angle is perpendicular to the paper. Then Panel a) shows the synchrotron polarized flux and polarization angle. Since the emission region is optically thin, Patch B will Compton scatter photons from all patches including itself. The SSC polarization with seed photons from Patch B itself simply reduces to the one-zone model and is well known in previous works \citep{Bonometto1973,Krawczynski2012,ZHC2013}. For the SSC with seed photons from other patches, say Patch A, the polarization depends on the scattering plane and the magnetic field in Patch A. Here we follow the convention in \citet{Bonometto1973}. For the Compton scattering of a beam of synchrotron seed photons, the polarization degree is given by
\begin{equation}
\begin{aligned}
\Pi_{ssc} & = \frac{P_{\perp}-P_{\parallel}}{P_{\perp}+P_{\parallel}} \\
 & = \frac{(Z_{\mathbf{\hat{e}'_\perp}}-Z_{\mathbf{\hat{e}'_\parallel}})(\Sigma_1+\Sigma_2)}{\Sigma_1+3\Sigma_2}
\end{aligned} ~~.
\end{equation}
Here
\begin{equation}
Z_{\mathbf{\hat{e}'}}=\left(\mathbf{\hat{e}}\cdot \mathbf{\hat{e}'}+\frac{(\mathbf{\hat{k}}\cdot  \mathbf{\hat{e}}')(\mathbf{\hat{k}'}\cdot  \mathbf{\hat{e}})}{1-\mathbf{\hat{k}}\cdot\mathbf{\hat{k}'}}\right)^2 ~~,
\end{equation}
where $\mathbf{\hat{e}}$, $\mathbf{\hat{k}}$ and $\mathbf{\hat{e}'}$, $\mathbf{\hat{k}'}$ are the unit polarization and direction vectors for the seed and scattered photons, respectively, and
\begin{equation}
\begin{aligned}
\Sigma_1 & =\int^{\beta_2}_{\beta_1}dx \, \frac{n(\gamma)}{\gamma^2}(x^2-\frac{1}{x^2}+2) \\
\Sigma_2 & =\int^{\beta_2}_{\beta_1}dx \, \frac{n(\gamma)}{\gamma^2}(x^2+\frac{1}{x^2}-2)
\end{aligned}~~,
\end{equation}
where $x=\gamma_{min}/\gamma$, $\beta_1=\min(\gamma_{min}/\gamma_2, 1)$, $\beta_2=\min(\gamma_{min}/\gamma_1, 1)$ for an electron distribution $n(\gamma)$ given in the range from $\gamma_1$ to $\gamma_2$, $\gamma_{min}$ is given by
\begin{equation}
\gamma_{min}=\sqrt{\frac{\epsilon'}{2\epsilon(1-\mathbf{\hat{k}}\cdot\mathbf{\hat{k}'})}}~~,
\end{equation}
and $\epsilon$, $\epsilon'$ are the seed and scattered photon energy normalized by the electron rest energy $m_ec^2$. \citet{Bonometto1973} defines $\mathbf{\hat{e}'_{\perp}}$ and $\mathbf{\hat{e}'_{\parallel}}$ as the unit vectors that are perpendicular and parallel to, respectively, the direction of the magnetic field $\mathbf{B}$ that produces the synchrotron seed photons projected onto the plane of the sky of the scattered photon, i.e., $\mathbf{\hat{e}'_{\perp}}$ is the unit vector along $\mathbf{B}\times \mathbf{\hat{k}'}$ and $\mathbf{\hat{e}'_{\parallel}}$ is the unit vector along $\mathbf{B}\times \mathbf{\hat{k}'} \times \mathbf{\hat{k}'}$. In this way, positive and negative values of $\Pi_{ssc}$ refer to polarization in the $\mathbf{\hat{e}'_{\perp}}$ and $\mathbf{\hat{e}'_{\parallel}}$ directions, respectively. It is clear that for a specific electron distribution $n(\gamma)$, $\Sigma_1$ and $\Sigma_2$ are effectively constants. The SSC polarization is determined by $Z_{\mathbf{\hat{e}'_\perp}}-Z_{\mathbf{\hat{e}'_\parallel}}=1-2Z_{\mathbf{\hat{e}'_\parallel}}$. For the scattering of seed photons from a specific patch to Patch B for instance, $\mathbf{\hat{k}}$ is fixed, and $\mathbf{\hat{k}'}$ is fixed by the viewing angle. But $\mathbf{\hat{e}}$ and $\mathbf{\hat{e}'}$ are random if the magnetic field $\mathbf{B}$ is random. Thus $1-2Z_{\mathbf{\hat{e}'_\parallel}}$ is a random value between $-1$ and $1$. If we take the absolute value, the SSC polarization degree is a random factor between 0 and 1 multiplied by a constant $(\Sigma_1+\Sigma_2)/(\Sigma_1+3\Sigma_2)$, and the polarization angle is random as well.

However, the polarized flux that contributes to the total polarized SSC flux from Patch B depends on the seed photon density as well. For a specific patch at a distance $a$ from Patch B, the seed photon density is proportional to $a^{-2}$. It is therefore tentative to consider the polarized flux from Patch B itself and very nearby patches, since their seed photon density dominates. However, the number of patches at a distance $a$ from Patch B is $N(a)\propto a^2$ as long as $a$ is smaller than the distance from Patch B to the edge of the blazar zone. If all patches have the same luminosity, then the flux contribution from patches at an arbitrary distance $a$ from Patch B can be considered as a constant $F_0$. But the total polarized flux at a distance $a$ is a different story. As mentioned before, the polarization angle for scattering of seed photons from an arbitrary patch is random due to the random magnetic field orientation. If we again make the assumption in deriving the ordering factor for synchrotron polarization that the flux and polarization degree are the same for seed photons from any patch (in practice, the polarization degree scales with a random number $|1-2Z_{\mathbf{\hat{e}'_\parallel}}|$),
then from Equation \ref{eq:pol} the total polarization degree for seed photons from a sphere of radius $a$ is given by
\begin{equation}
\label{PiSSCtot}
\Pi_{ssc,tot}(a)=\frac{1}{\sqrt{N(a)}}\Pi_{ssc}\propto \frac{1}{a}\Pi_{ssc}~~,
\end{equation}
and the total polarization angle is obviously random. If we define $a=1$ for Patch B itself, $a=2$ for the closest patch (Patch A in Figure \ref{fig:sscsketch}, for instance), etc., then, since $N$ is a finite number, there will be $a_{max}\propto N^{1/3}$ spheres that contribute to the total flux and polarized flux for SSC by Patch B. We can approximate the total flux as
\begin{equation}
F_{tot}=\sum_{a=1}^{a_{max}}F_0=a_{max}F_0 ~~,
\end{equation}
and the total polarized flux is given by
\begin{equation}
\begin{aligned}
F_{pol} & =\sqrt{Q_{tot}^2+U_{tot}^2} \\
Q_{tot} & =\sum_{a=1}^{a_{max}}\Pi_{ssc,tot}(a)F_0 \cos(2\mathcal{R}(PA)) \\
U_{tot} & =\sum_{a=1}^{a_{max}}\Pi_{ssc,tot}(a)F_0 \sin(2\mathcal{R}(PA)) 
\end{aligned} ~~,
\end{equation}
where $\mathcal{R}(PA)$ is a random polarization angle.
$\Pi_{ssc,tot}(a)$ is given by Eq. \ref{PiSSCtot}, thus we have
\begin{equation}
F_{pol}\propto\Pi_{ssc}F_0 \sqrt{\sum_{a=1}^{a_{max}} \frac{1}{a}(\cos^2(2\mathcal{R}(PA))+\sin^2(2\mathcal{R}(PA)))} ~~.
\end{equation}
The term inside the square root is a random walk in the Stokes $Q$-$U$ plane with a step size that decreases with $a$. It does not have a simple analytical solution, but clearly the first step $a=1$, i.e., the polarized flux from Patch B itself dominates. If we make the approximation that we only consider the polarized flux from Patch B itself, then we have the SSC polarization degree from Patch B as
\begin{equation}
\Pi_0=\frac{F_{pol}}{F_{tot}}=\frac{\Pi_{ssc}F_0}{N^{1/3}F_0}=\frac{1}{N^{1/3}}\Pi_{ssc} ~~.
\label{eq:sscone}
\end{equation}
The factor $\frac{1}{N^{1/3}}$ is derived following the same approximations and arguments as in deriving the ordering factor of synchrotron polarization, and we call this factor the secondary depolarization factor.

Panel b) in Figure \ref{fig:sscsketch} illustrates the SSC polarized flux with seed photons contributed by each patch in the emission region. The secondary depolarization factor should apply to all patches, and Panel c) shows the same illustration as Panel b), but for Patch C. It is clear from both panels that the polarized SSC flux is dominated by the SSC of seed photons from the patch itself. Panel d) then shows the total polarized SSC flux from each patch. The polarization degrees are comparable, and the polarization angles are random since the magnetic field in each patch is random. If the emission region is optically thin to the SSC flux, which is generally true for blazar emission below TeV, the total SSC polarization degree can be obtained by inserting Equation \ref{eq:sscone} into Equation \ref{eq:pol}. If we further assume that the magnetic field is perfectly ordered in each patch, then previous one-zone models show that $\Pi_{ssc,max}\sim 50\% \Pi_{syn,max}$, thus we have
\begin{equation}
\Pi_{ssc,tot}=\frac{1}{\sqrt{N}}\frac{1}{N^{1/3}}\Pi_{ssc,max}\sim 50\%\frac{1}{N^{5/6}}\Pi_{syn,max} ~~.
\label{eq:ssc}
\end{equation}
Comparing it with Equation \ref{eq:syn}, we can easily see that the differences are the secondary depolarization factor and the $\sim 50\%$ depolarization intrinsically from the SSC. Although the $N^{1/3}$ factor may appear small, we note that typically the blazar optical synchrotron polarization degree is about $\lesssim 10\%$. Based on Equation \ref{eq:syn}, this implies $N\gtrsim 50$ if $\Pi_{syn,max}\sim 70\%$. Thus the SSC polarization degree can only reach about $1\sim 2\%$, mostly undetectable with current and 
planned X-ray and $\gamma$-ray polarimeters (see Section \ref{sec:obs}).

As mentioned above, the double depolarization in Equation \ref{eq:ssc} suffers from the same four issues discussed in the case of the ordering factor of synchrotron polarization. However, we note that the third issue, temporal evolution, can behave differently for SSC compared to synchrotron. Since the SSC polarization depends on the polarization of the seed synchrotron photons, it is expected to be as highly variable as the synchrotron polarization. However, the synchrotron and SSC polarization variations do not necessarily correlate. This is due to the extra internal light crossing time delay affecting SSC. Both synchrotron and SSC suffer from the external light crossing time delay, that is the light crossing time from a specific patch to the observer. However, as illustrated in Panel d) in Figure \ref{fig:sscsketch}, the seed photons from Patch A arrive at different times in Patches B and C, which is referred to as the internal light crossing time delay. In an inhomogeneous and evolving blazar zone, it is possible that the seed photons from Patch A contribute significantly to either the photon density or the SSC polarized flux in Patches B and/or C. Then the internal light crossing time of SSC will exhibit a time-delayed view of the physical evolution in Patch A compared to synchrotron.

Unfortunately, it is very computationally expensive to test the above double depolarization effect. \citet{Peirson2019} finds that their code will take 16 CPUs running more than 4 hours for an emission region of just $N_{tot}=N=37$ cells, and the computational cost scales with $N^2_{tot}$. Aforementioned, the derivation of both the ordering factor of synchrotron polarization and double depolarization of SSC is based on statistical quantities, requiring that $N_{tot}\gg N\gg 1$. As a result, a reasonable and comprehensive test of the double depolarization effect will need at least a factor of a few tens more cells than that used in \citet{Peirson2019}, which translates to at least a thousand times more computing resources. That is far beyond the scope of this paper. Nonetheless, it is interesting to note that \citet{Peirson2019} suggests a relation similar to Equation \ref{eq:ssc}: $\Pi_{ssc}\propto N_{eff}^{-p}N^{-1/2}$, where $N_{eff}$ is one of their model parameter that describes the number of cells contributing to half of the total flux. The paper shows that $p\sim 0.25$, but its value can change based on the seed photon's angular polarization distribution in their model. This $N_{eff}^{-0.25}$ has a very similar physical interpretation as the secondary depolarization factor mentioned above, except that we consider a generic blazar zone consisting of patches with uncorrelated magnetic fields.

\section{Energy Stratification \label{sec:hadronic}}

This section describes the energy stratification effect that can affect the hadronic polarization in an emission region with inhomogeneous magnetic fields. We first show that the energy stratification can lead to polarization differences in the synchrotron component that scale with $\gamma_e^{3(1-\alpha)/4}$ if the acceleration of low- and high-energy electrons is co-spatial, where $\alpha$ is the energy-dependency of the diffusion coefficient. If the hadronic emission flares contemporaneously with the synchrotron component, the polarization degree scales with $t_{var}^{-3/2}$, where $t_{var}$ is the variability time scale. Another possibility is that the hadronic polarization degree may be correlated with the radio polarization if protons diffuse and occupy the large-scale jet.

\subsection{Energy Stratification in Primary Synchrotron Component}

\begin{figure}
\centering
\includegraphics[width=0.99\linewidth]{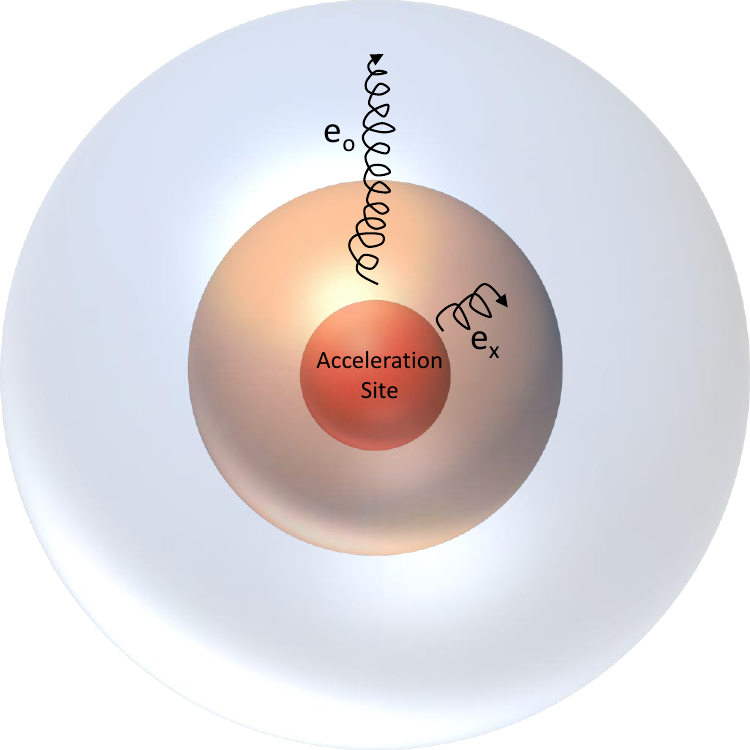}
\caption{A sketch of the energy stratification due to diffusion for HSP blazars. Due to the shorter cooling time scale of high-energy electrons that synchrotron emit in the X-ray band, it is possible that they can diffuse a smaller radius from the acceleration site compared to those that synchrotron emit in the optical band. Thus the X-ray polarization degree can be higher than that of the optical band.}
\label{fig:diffusionsketch}
\end{figure}

IXPE observations of bright HSPs, such as Mrk~421 and Mrk~501, yield much higher X-ray polarization degrees than those in the optical and radio bands \citep{Liodakis2022,DiGesu2022}. This is expected under the energy stratification model. Consider a blazar zone that consists of $N$ patches of uncorrelated magnetic fields. Assume that all electrons are accelerated co-spatially, they can diffuse and/or advect out of the acceleration site (Figure \ref{fig:diffusionsketch} shows an example for the diffusion model). Due to the radiative cooling, higher-energy electrons that emit synchrotron radiation in the X-ray bands will travel a shorter distance than lower-energy electrons that emit in the optical bands, thus they are spread out over a smaller region. Such phenomena are shown in many radiative PIC simulations of shock, reconnection, and turbulence \citep[e.g.,][]{ZHC2020,ZHC2023}. The ratio of the X-ray to the optical polarization degree, according to Equation \ref{eq:syn}, depends on the ratio of the number of patches occupied by X-ray emitting electrons over that occupied by optical emitting electrons, i.e.,
\begin{equation}
\frac{\Pi_X}{\Pi_o}=\sqrt{\frac{N_o}{N_X}}~~.
\label{eq:stratification}
\end{equation}
The spatial diffusion coefficient of the electrons is $D(\gamma_e)\propto \gamma_e^\alpha$, where $\gamma_e$ is the Lorentz factor of the nonthermal electron and $\alpha$ describes the energy-dependency of the diffusion coefficient, which can range from $1/3$ (motivated by the Kolmogorov turbulence model) to $1$ (Bohm approximation). Since the radiative cooling time scale of both synchrotron and inverse Compton scattering (in the Thomson regime) is anti-proportional to $\gamma_e$, the cooling time scale $\tau\propto \gamma_e^{-1}$. Hence, the distance that a nonthermal electron can travel is given by
\begin{equation}
r=\sqrt{D\tau}\propto \gamma_e^{\frac{\alpha-1}{2}}~~.
\end{equation}
Since $N\propto r^3$, we can find that the X-ray polarization degree is higher than the optical by
\begin{equation}
\frac{\Pi_X}{\Pi_o}=\sqrt{\frac{r_o^3}{r_X^3}}=(\frac{\gamma_X}{\gamma_o})^{\frac{3(1-\alpha)}{4}}~~.
\end{equation}
where $\gamma_X$ and $\gamma_o$ are the energies of electrons emitting synchrotron radiation in the X-ray and optical bands, respectively. Since the synchrotron critical frequency is proportional to $\gamma_e^2$, if the magnetic field strength is similar throughout the emission region, then $\gamma_X/\gamma_o\propto \sqrt{1000}\gtrsim 30$. For $1/3\le \alpha \le 1$, the above ratio ranges from $\sim 5$ to 1, consistent with the observations. Note that the above calculation only considers diffusion. Other spatial transport mechanisms, such as advection, can have different electron energy dependency. For instance, the stratification along the jet propagation direction in \citet{Liodakis2022} and \citet{DiGesu2022} is effectively an advection model.

\subsection{Energy Stratification in the High-Energy Component}

Energy stratification should apply to hadronic models as well. However, the above calculation is probably irrelevant in most cases. This is because under the proton synchrotron model, the synchrotron cooling dominates for both electrons and protons, and the cooling time scales with $\tau\propto m^3/\gamma$. Then the ratio of diffusion radius of protons over that of electrons becomes $r_p/r_e\propto (m_p/m_e)^{(\alpha+3)/2}(\gamma_p/\gamma_e)^{(\alpha-1)/2}$. Since the synchrotron critical frequency scales with $\gamma^2/m$, considering that the MeV band is produced through proton synchrotron and the optical band through electron synchrotron, then $\gamma_p^2/\gamma_e^2=10^6m_p/m_e$. Thus we have $\gamma_p/\gamma_e\sim 4\times 10^4$, and $r_p/r_e$ ranges from $10^4$ to $4\times 10^6$ for $\alpha$ between 1/3 and 1. It can be shown similarly that typical hadronic blazar parameters yield $r_p/r_e$ in roughly the same range or even larger for other hadronic processes. Such a large ratio should generally result in very different variability time scales of low- and high-energy spectral components, which are not supported by observations. This suggests that the sizes of the regions occupied by protons and electrons are not determined by the diffusion radius. For example, physical conditions for significant hadronic processes such as target photon density and magnetic field strengths may be satisfied only in a blazar zone of relatively small size ($R\sim 10^{16}$ cm for instance) that is much smaller than $r_p$.

The other way to constrain the sizes of regions occupied by protons and electrons is using the variability time scales. When the blazar exhibits multi-wavelength flares, the size of the flaring region can be well constrained by the causality relation. The region occupied by electrons that emit synchrotron radiation in the optical band, for instance, is simply $R_o\le c\delta t_{var,o}$, where $\delta$ is the relativistic Doppler factor and $t_{var,o}$ is the observed variability time scale in the optical band. Similarly, if the X-ray and/or $\gamma$-ray emissions are of hadronic origin, $R_{had}\le c\delta t_{var,had}$. Previous works show that symmetric light curves, i.e., the rising and falling slopes of the light curves are comparable, suggest
that the light crossing time determines the shape of light curves \citep{Chen2011,ZHC2015}. In this case, $R=c\delta t_{var}$, and the hadronic polarization degree can be estimated by
\begin{equation}
\frac{\Pi_{had}}{\Pi_o}=\sqrt{\frac{N_o}{N_{had}}}=(\frac{R_o}{R_{had}})^{\frac{3}{2}}=(\frac{t_{var,o}}{t_{var,had}})^{\frac{3}{2}}~~.
\end{equation}
Observations often find that the optical and the high-energy bands flare simultaneously \citep[e.g.,][]{Rani2013,Liodakis2018,Liodakis2019}. If both bands show symmetric light curves and $t_{var,o}/t_{var,had}\sim 1$, then the multi-zone prediction is consistent with the one-zone prediction, $\Pi_{had}\sim \Pi_o$. We note that these estimates for flaring and quiescent states refer to the intrinsic hadronic polarization degree. Some lepto-hadronic spectral models have a considerable leptonic contribution in the high-energy component \citep{Boettcher2013,Cerruti2015}. In this case, the predicted X-ray and MeV polarization degree can be lower depending on how much the hadronic emission contributes.

If the blazar is in the quiescent state or the light curve of the flare is clearly asymmetric, it is not possible to constrain the hadronic polarization without fully simulating the jet evolution and particle dynamics. One possibility is, however, worth mentioning: Radio observations often find large-scale magnetic structures in some blazar jets, whose shape can remain relatively steady over years \citep[e.g.,][]{Jorstad2017}. As mentioned before, protons can have a very large diffusion radius. As an example, consider a PeV proton that experiences Bohm diffusion and synchrotron cooling. In a magnetic field of 0.1~G the diffusion radius is $r_p\sim 10^{23}$ cm. Note that the photomeson and Bethe-Heitler processes may cool protons faster than synchrotron if the target photon density is high, but given that $r_p\propto \sqrt{\tau}$, $r_p$ is unlikely many orders of magnitude lower than this estimate. It is reasonable to assume that protons may occupy the large-scale jet that emits in radio if hadronic processes can contribute considerably to the quiescent blazar emission. In this situation, if the exposure time is much longer than the typical variability time scales, for example, $\gtrsim 1$ year, hadronic models may yield a non-trivial polarization degree comparable to the radio polarization, which originates from the relatively steady large-scale magnetic structure.

Finally, we note that the energy stratification can have an impact on the SSC polarization as well, since the observed X-ray and $\gamma$-ray emission can result from a range of nonthermal electron and seed photon energies in SSC. This effect, along with the internal light crossing time delays, can smoothen the SSC flux variability as shown in \citet{Chen2011}. Thus we expect that it may reduce the SSC polarization as well. A fully 3D SSC radiation transfer simulation is necessary to evaluate this effect.

\section{Implications for Observations \label{sec:obs}}

This section presents a discussion of the detectability of leptonic and hadronic polarization in the high-energy spectral component with IXPE and future high-energy polarimeters, based on the multi-zone effects described in the previous sections. We find that the leptonic polarization may be detectable by IXPE within a 500~ksec exposure time for some of the brightest LSP/ISP blazars, if the intrinsic synchrotron optical polarization degree remains $\gtrsim 25\%$ and the polarization angle is steady as well. But the hadronic polarization degree can be easily detectable within a 100~ksec exposure if the optical and high-energy bands flare simultaneously. For future MeV polarimeters such as COSI and AMEGO-X, due to the low photon flux in the MeV band, even the detection of hadronic polarization signatures can be challenging unless the blazar MeV flux reaches $>0.1$ Crab units. Alternatively, if MeV polarimeters can monitor blazars and achieve a significant non-zero polarization after $\gtrsim 1$ year exposure, it would be strong evidence for hadronic models.

\begin{table}
\centering
\begin{tabular}{|c|c|c|c|c|c|c|} \hline
$\Pi_o$      & 5\%    & 10\%   & 15\%   & 20\%   & 25\%   & 40\%   \\ \hline
$\Pi_{SSC}$  & 0.43\% & 1.4\%  & 2.7\%  & 4.3\%  & 6.3\%  & 14\%   \\ \hline
\end{tabular}
\caption{The SSC polarization degree $\Pi_{SSC}$ for different optical synchrotron polarization degree. We assume a maximal synchrotron polarization degree of $\Pi_{syn,max}=70\%$.}
\label{tab:ssc}
\end{table}

Table \ref{tab:ssc} shows the SSC polarization degrees for different intrinsic synchrotron optical polarization degrees, i.e., not contaminated by other weakly polarized emission contributions such as the accretion disk and the host galaxy. The actual leptonic polarization degree may be lower if EC contributes significantly to the high-energy bands, but here we consider the SSC polarization to maximize detectability. To explore the limits of the detectability of SSC polarization for IXPE, we use the X-ray flux modeling in \cite{Liodakis2019-II} to identify average and flaring states for four of the brightest LSP/ISP sources, namely 3C~273 (LSP), 3C~454.3 (LSP), S5~0716+71 (ISP), BL Lacertae (LSP/ISP). \cite{Liodakis2019-II} used the Bayesian methodology developed in \cite{Liodakis2017} to fit the long-term X-ray light curves from Swift and RXTE with a single-Gaussian and a double-Gaussian model. The best-fit model was then selected according to the Bayesian Information Criterion. Using the long-term flux for different states we estimate the Minimum Detectable Polarization at the 99\% confidence interval (MDP99). The MDP99 defines the polarization degree level that any measured polarization signal has a less than 1\% probability of being due to random fluctuations of the background. Hence, any measurement above the MDP99 can be considered intrinsic to the source.

\begin{figure*}
\centering
\includegraphics[width=0.99\linewidth]{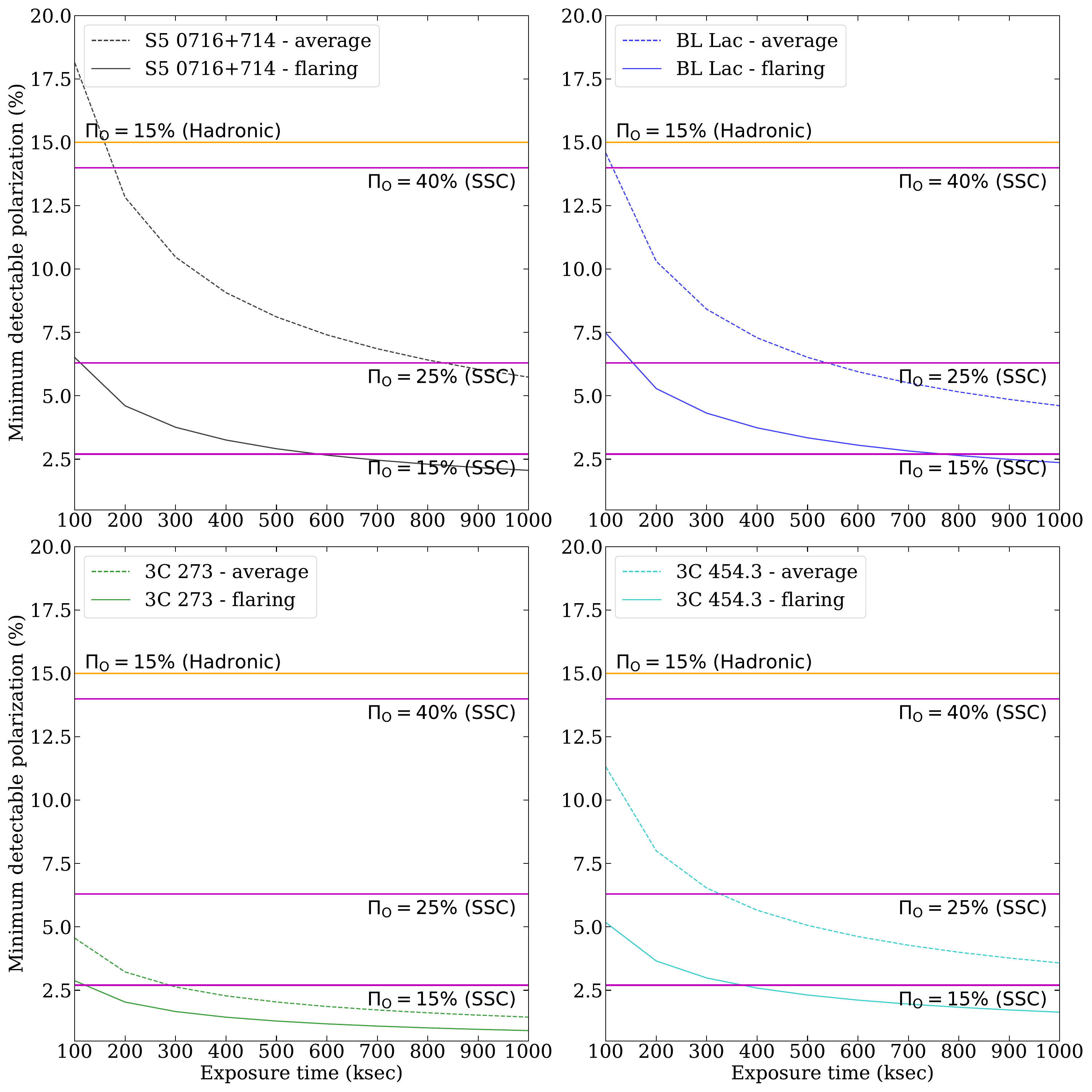}
\caption{MDP99 versus exposure time for four of the brightest LSP/ISP blazars. Each panel shows the estimate for the average and flaring state of each blazar. In all panels the magenta lines show the level of SSC polarization for 40~\%, 25~\%, and 15~\% optical polarization (Table \ref{tab:ssc}), and the orange lines show the hadronic polarization for 15~\% optical polarization during simultaneous multi-wavelength flares.}
\label{fig:ixpe_mdp}
\end{figure*}

Figure \ref{fig:ixpe_mdp} shows the expected MDP99 versus exposure time for the aforementioned blazars. It is obvious that within a reasonable exposure time ($\rm <500~ksec$, given that blazar variability can lead to depolarization), the optical polarization needs to be at least $\gtrsim 15\%$ for the flaring states, and $\gtrsim 25\%$ for the average states of these four blazars. Given the average efficiency of IXPE observations related to the spacecraft orbit, a 500~ksec observation would roughly correspond to 9 days. While such high levels of optical polarization are not atypical of blazars, especially for 3C~279 and 3C~454.3, which have often been found to be $>30\%$ polarized \citep{Liodakis2020-II,Blinov2021,Liodakis2022-II}, it is unusual that both the optical polarization degree and angle are stable for an observational time longer than a week. As mentioned before, if significant variations in polarization occur, both the synchrotron and SSC polarization degrees can be averaged out due to polarization angle changes. 3C~273 shows the best prospects for detecting the SSC polarization. Unfortunately, the source is notoriously unpolarized, showing only 1 -- 2~\% polarization in the infrared/optical due to the strong emission from the accretion disk \citep[e.g.,][]{Marshall2023}, thus its intrinsic synchrotron polarization is not well constrained. Additionally, LSP and ISP blazars often have a considerable EC component, which can significantly lower the overall leptonic polarization. The enhanced X-ray Timing and Polarimetry mission (eXTP, \citealp{extp2019}) is expected to reduce the MDP99 for the same exposure time by a factor of 2 compared to IXPE \citep{Peirson2022}. This will certainly improve the detectability of the leptonic polarization. However, it will still require flaring blazars with a steady $>10\%$ optical polarization to achieve a detection in $\rm <500~ksec$ exposures. On the other hand, the hadronic polarization degree can be comparable to the optical polarization if the optical and high-energy bands flare simultaneously. This can be easily detectable with IXPE in $\rm <100~ksec$ if the optical polarization degree is $\gtrsim 15\%$, which is quite common during flares. \citet{ZHC2016b} shows that the hadronic polarization can be stable during relatively short flares, boosting the chance that IXPE detects hadronic polarization signatures.

It is very challenging to detect polarization in the MeV band due to the much lower photon flux. Future MeV polarimeters are unlikely to detect any leptonic polarization. COSI may detect hadronic polarization in a couple of weeks if the MeV and optical bands flare simultaneously and the MeV flux reaches $\gtrsim 0.5$ Crab units. In such a strong flare, COSI can confirm hadronic contributions if the hadronic polarization degree reaches $\gtrsim 20\%$. COSI's ongoing data challenge 3 will provide clearer estimates on the polarization capability. More powerful polarimeter concepts such as AMEGO-X will be able to detect hadronic polarization at $\sim 15\%$ in about one week if the MeV flux reaches $\gtrsim 0.1$ Crab units. On the other hand, owing to their large field of view, COSI and AMEGO-X can monitor blazars, achieving $\sim 10\%$ or even lower MDP99 after an exposure time of $\sim 1$ year. If polarization is detected on such a long time scale, it can be strong evidence for hadronic models.

\begin{acknowledgements}
We thank the anonymous referee for very constructive comments. HZ is supported by NASA under award No. 80GSFC21M0002. HZ is supported by Fermi Guest Investigator Program Cycle 16, award No. 22-FERMI22-0015. IL was supported by the NASA Postdoctoral Program at the Marshall Space Flight Center, administered by Oak Ridge Associated Universities under contract with NASA.
The work of MB is supported in part by the Department of Science and Innovation and the National Research Foundation of South Africa through a grant in support of the South African Gamma-Ray Astronomy Programme (SA-GAMMA). 
\end{acknowledgements}

\bibliography{Blazar}{}

\begin{thebibliography}{}
\expandafter\ifx\csname natexlab\endcsname\relax\def\natexlab#1{#1}\fi
\providecommand{\url}[1]{\href{#1}{#1}}
\providecommand{\dodoi}[1]{doi:~\href{http://doi.org/#1}{\nolinkurl{#1}}}
\providecommand{\doeprint}[1]{\href{http://ascl.net/#1}{\nolinkurl{http://ascl.net/#1}}}
\providecommand{\doarXiv}[1]{\href{https://arxiv.org/abs/#1}{\nolinkurl{https://arxiv.org/abs/#1}}}

\bibitem[{{Abdollahi} {et~al.}(2020){Abdollahi}, {Acero}, {Ackermann}, {Ajello}, {Atwood}, {Axelsson}, {Baldini}, {Ballet}, {Barbiellini}, {Bastieri}, {Becerra Gonzalez}, {Bellazzini}, {Berretta}, {Bissaldi}, {Blandford}, {Bloom}, {Bonino}, {Bottacini}, {Brandt}, {Bregeon}, {Bruel}, {Buehler}, {Burnett}, {Buson}, {Cameron}, {Caputo}, {Caraveo}, {Casandjian}, {Castro}, {Cavazzuti}, {Charles}, {Chaty}, {Chen}, {Cheung}, {Chiaro}, {Ciprini}, {Cohen-Tanugi}, {Cominsky}, {Coronado-Bl{\'a}zquez}, {Costantin}, {Cuoco}, {Cutini}, {D'Ammando}, {DeKlotz}, {de la Torre Luque}, {de Palma}, {Desai}, {Digel}, {Di Lalla}, {Di Mauro}, {Di Venere}, {Dom{\'\i}nguez}, {Dumora}, {Fana Dirirsa}, {Fegan}, {Ferrara}, {Franckowiak}, {Fukazawa}, {Funk}, {Fusco}, {Gargano}, {Gasparrini}, {Giglietto}, {Giommi}, {Giordano}, {Giroletti}, {Glanzman}, {Green}, {Grenier}, {Griffin}, {Grondin}, {Grove}, {Guiriec}, {Harding}, {Hayashi}, {Hays}, {Hewitt}, {Horan}, {J{\'o}hannesson}, {Johnson}, {Kamae}, {Kerr}, {Kocevski}, {Kovac'evic'},
  {Kuss}, {Landriu}, {Larsson}, {Latronico}, {Lemoine-Goumard}, {Li}, {Liodakis}, {Longo}, {Loparco}, {Lott}, {Lovellette}, {Lubrano}, {Madejski}, {Maldera}, {Malyshev}, {Manfreda}, {Marchesini}, {Marcotulli}, {Mart{\'\i}-Devesa}, {Martin}, {Massaro}, {Mazziotta}, {McEnery}, {Mereu}, {Meyer}, {Michelson}, {Mirabal}, {Mizuno}, {Monzani}, {Morselli}, {Moskalenko}, {Negro}, {Nuss}, {Ojha}, {Omodei}, {Orienti}, {Orlando}, {Ormes}, {Palatiello}, {Paliya}, {Paneque}, {Pei}, {Pe{\~n}a-Herazo}, {Perkins}, {Persic}, {Pesce-Rollins}, {Petrosian}, {Petrov}, {Piron}, {Poon}, {Porter}, {Principe}, {Rain{\`o}}, {Rando}, {Razzano}, {Razzaque}, {Reimer}, {Reimer}, {Remy}, {Reposeur}, {Romani}, {Saz Parkinson}, {Schinzel}, {Serini}, {Sgr{\`o}}, {Siskind}, {Smith}, {Spandre}, {Spinelli}, {Strong}, {Suson}, {Tajima}, {Takahashi}, {Tak}, {Thayer}, {Thompson}, {Tibaldo}, {Torres}, {Torresi}, {Valverde}, {Van Klaveren}, {van Zyl}, {Wood}, {Yassine}, \& {Zaharijas}}]{Fermi2020}
{Abdollahi}, S., {Acero}, F., {Ackermann}, M., {et~al.} 2020, \apjs, 247, 33, \dodoi{10.3847/1538-4365/ab6bcb}

\bibitem[{{Ackermann} {et~al.}(2016){Ackermann}, {Anantua}, {Asano}, {Baldini}, {Barbiellini}, {Bastieri}, {Becerra Gonzalez}, {Bellazzini}, {Bissaldi}, {Blandford}, {Bloom}, {Bonino}, {Bottacini}, {Bruel}, {Buehler}, {Caliandro}, {Cameron}, {Caragiulo}, {Caraveo}, {Cavazzuti}, {Cecchi}, {Cheung}, {Chiang}, {Chiaro}, {Ciprini}, {Cohen-Tanugi}, {Costanza}, {Cutini}, {D'Ammando}, {de Palma}, {Desiante}, {Digel}, {Di Lalla}, {Di Mauro}, {Di Venere}, {Drell}, {Favuzzi}, {Fegan}, {Ferrara}, {Fukazawa}, {Funk}, {Fusco}, {Gargano}, {Gasparrini}, {Giglietto}, {Giordano}, {Giroletti}, {Grenier}, {Guillemot}, {Guiriec}, {Hayashida}, {Hays}, {Horan}, {J{\'o}hannesson}, {Kensei}, {Kocevski}, {Kuss}, {La Mura}, {Larsson}, {Latronico}, {Li}, {Longo}, {Loparco}, {Lott}, {Lovellette}, {Lubrano}, {Madejski}, {Magill}, {Maldera}, {Manfreda}, {Mayer}, {Mazziotta}, {Michelson}, {Mirabal}, {Mizuno}, {Monzani}, {Morselli}, {Moskalenko}, {Nalewajko}, {Negro}, {Nuss}, {Ohsugi}, {Orlando}, {Paneque}, {Perkins}, {Pesce-Rollins},
  {Piron}, {Pivato}, {Porter}, {Principe}, {Rando}, {Razzano}, {Razzaque}, {Reimer}, {Scargle}, {Sgr{\`o}}, {Sikora}, {Simone}, {Siskind}, {Spada}, {Spinelli}, {Stawarz}, {Thayer}, {Thompson}, {Torres}, {Troja}, {Uchiyama}, {Yuan}, \& {Zimmer}}]{Ackermann2016}
{Ackermann}, M., {Anantua}, R., {Asano}, K., {et~al.} 2016, \apjl, 824, L20, \dodoi{10.3847/2041-8205/824/2/L20}

\bibitem[{{Albert} {et~al.}(2007){Albert}, {Aliu}, {Anderhub}, {Antoranz}, {Armada}, {Baixeras}, {Barrio}, {Bartko}, {Bastieri}, {Becker}, {Bednarek}, {Berger}, {Bigongiari}, {Biland}, {Bock}, {Bordas}, {Bosch-Ramon}, {Bretz}, {Britvitch}, {Camara}, {Carmona}, {Chilingarian}, {Coarasa}, {Commichau}, {Contreras}, {Cortina}, {Costado}, {Curtef}, {Danielyan}, {Dazzi}, {De Angelis}, {Delgado}, {de los Reyes}, {De Lotto}, {Domingo-Santamar{\'\i}a}, {Dorner}, {Doro}, {Errando}, {Fagiolini}, {Ferenc}, {Fern{\'a}ndez}, {Firpo}, {Flix}, {Fonseca}, {Font}, {Fuchs}, {Galante}, {Garc{\'\i}a-L{\'o}pez}, {Garczarczyk}, {Gaug}, {Giller}, {Goebel}, {Hakobyan}, {Hayashida}, {Hengstebeck}, {Herrero}, {H{\"o}hne}, {Hose}, {Hrupec}, {Hsu}, {Jacon}, {Jogler}, {Kosyra}, {Kranich}, {Kritzer}, {Laille}, {Lindfors}, {Lombardi}, {Longo}, {L{\'o}pez}, {L{\'o}pez}, {Lorenz}, {Majumdar}, {Maneva}, {Mannheim}, {Mansutti}, {Mariotti}, {Mart{\'\i}nez}, {Mazin}, {Merck}, {Meucci}, {Meyer}, {Miranda}, {Mirzoyan}, {Mizobuchi}, {Moralejo},
  {Nieto}, {Nilsson}, {Ninkovic}, {O{\~n}a-Wilhelmi}, {Otte}, {Oya}, {Paneque}, {Panniello}, {Paoletti}, {Paredes}, {Pasanen}, {Pascoli}, {Pauss}, {Pegna}, {Persic}, {Peruzzo}, {Piccioli}, {Prandini}, {Puchades}, {Raymers}, {Rhode}, {Rib{\'o}}, {Rico}, {Rissi}, {Robert}, {R{\"u}gamer}, {Saggion}, {Saito}, {S{\'a}nchez}, {Sartori}, {Scalzotto}, {Scapin}, {Schmitt}, {Schweizer}, {Shayduk}, {Shinozaki}, {Shore}, {Sidro}, {Sillanp{\"a}{\"a}}, {Sobczynska}, {Stamerra}, {Stark}, {Takalo}, {Tavecchio}, {Temnikov}, {Tescaro}, {Teshima}, {Torres}, {Turini}, {Vankov}, {Vitale}, {Wagner}, {Wibig}, {Wittek}, {Zandanel}, {Zanin}, \& {Zapatero}}]{Albert2007}
{Albert}, J., {Aliu}, E., {Anderhub}, H., {et~al.} 2007, \apj, 669, 862, \dodoi{10.1086/521382}

\bibitem[{{Blinov} {et~al.}(2021){Blinov}, {Kiehlmann}, {Pavlidou}, {Panopoulou}, {Skalidis}, {Angelakis}, {Casadio}, {Einoder}, {Hovatta}, {Kokolakis}, {Kougentakis}, {Kus}, {Kylafis}, {Kyritsis}, {Lalakos}, {Liodakis}, {Maharana}, {Makrydopoulou}, {Mandarakas}, {Maragkakis}, {Myserlis}, {Papadakis}, {Paterakis}, {Pearson}, {Ramaprakash}, {Readhead}, {Reig}, {S{\l}owikowska}, {Tassis}, {Xexakis}, {{\.Z}ejmo}, \& {Zensus}}]{Blinov2021}
{Blinov}, D., {Kiehlmann}, S., {Pavlidou}, V., {et~al.} 2021, \mnras, 501, 3715, \dodoi{10.1093/mnras/staa3777}

\bibitem[{{Bonometto} \& {Saggion}(1973)}]{Bonometto1973}
{Bonometto}, S., \& {Saggion}, A. 1973, \aap, 23, 9

\bibitem[{{B{\"o}ttcher}(2019)}]{Boettcher2019}
{B{\"o}ttcher}, M. 2019, Galaxies, 7, 20, \dodoi{10.3390/galaxies7010020}

\bibitem[{{B{\"o}ttcher} {et~al.}(2013){B{\"o}ttcher}, {Reimer}, {Sweeney}, \& {Prakash}}]{Boettcher2013}
{B{\"o}ttcher}, M., {Reimer}, A., {Sweeney}, K., \& {Prakash}, A. 2013, \apj, 768, 54, \dodoi{10.1088/0004-637X/768/1/54}

\bibitem[{{Cerruti} {et~al.}(2019){Cerruti}, {Zech}, {Boisson}, {Emery}, {Inoue}, \& {Lenain}}]{Cerruti2019}
{Cerruti}, M., {Zech}, A., {Boisson}, C., {et~al.} 2019, \mnras, 483, L12, \dodoi{10.1093/mnrasl/sly210}

\bibitem[{{Cerruti} {et~al.}(2015){Cerruti}, {Zech}, {Boisson}, \& {Inoue}}]{Cerruti2015}
{Cerruti}, M., {Zech}, A., {Boisson}, C., \& {Inoue}, S. 2015, \mnras, 448, 910, \dodoi{10.1093/mnras/stu2691}

\bibitem[{{Chen} {et~al.}(2011){Chen}, {Fossati}, {Liang}, \& {B{\"o}ttcher}}]{Chen2011}
{Chen}, X., {Fossati}, G., {Liang}, E.~P., \& {B{\"o}ttcher}, M. 2011, \mnras, 416, 2368, \dodoi{10.1111/j.1365-2966.2011.19215.x}

\bibitem[{{Dermer} {et~al.}(1992){Dermer}, {Schlickeiser}, \& {Mastichiadis}}]{Dermer1992}
{Dermer}, C.~D., {Schlickeiser}, R., \& {Mastichiadis}, A. 1992, \aap, 256, L27

\bibitem[{{Di Gesu} {et~al.}(2022){Di Gesu}, {Donnarumma}, {Tavecchio}, {Agudo}, {Barnounin}, {Cibrario}, {Di Lalla}, {Di Marco}, {Escudero}, {Errando}, {Jorstad}, {Kim}, {Kouch}, {Liodakis}, {Lindfors}, {Madejski}, {Marshall}, {Marscher}, {Middei}, {Muleri}, {Myserlis}, {Negro}, {Omodei}, {Pacciani}, {Paggi}, {Perri}, {Puccetti}, {Antonelli}, {Bachetti}, {Baldini}, {Baumgartner}, {Bellazzini}, {Bianchi}, {Bongiorno}, {Bonino}, {Brez}, {Bucciantini}, {Capitanio}, {Castellano}, {Cavazzuti}, {Ciprini}, {Costa}, {De Rosa}, {Del Monte}, {Doroshenko}, {Dov{\v{c}}iak}, {Ehlert}, {Enoto}, {Evangelista}, {Fabiani}, {Ferrazzoli}, {Garcia}, {Gunji}, {Hayashida}, {Heyl}, {Iwakiri}, {Karas}, {Kitaguchi}, {Kolodziejczak}, {Krawczynski}, {La Monaca}, {Latronico}, {Maldera}, {Manfreda}, {Marin}, {Marinucci}, {Massaro}, {Matt}, {Mitsuishi}, {Mizuno}, {Ng}, {O'Dell}, {Oppedisano}, {Papitto}, {Pavlov}, {Peirson}, {Pesce-Rollins}, {Petrucci}, {Pilia}, {Possenti}, {Poutanen}, {Ramsey}, {Rankin}, {Ratheesh}, {Romani}, {Sgr{\`o}},
  {Slane}, {Soffitta}, {Spandre}, {Tamagawa}, {Taverna}, {Tawara}, {Tennant}, {Thomas}, {Tombesi}, {Trois}, {Tsygankov}, {Turolla}, {Vink}, {Weisskopf}, {Wu}, {Xie}, \& {Zane}}]{DiGesu2022}
{Di Gesu}, L., {Donnarumma}, I., {Tavecchio}, F., {et~al.} 2022, \apjl, 938, L7, \dodoi{10.3847/2041-8213/ac913a}

\bibitem[{{Di Gesu} {et~al.}(2023){Di Gesu}, {Marshall}, {Ehlert}, {Kim}, {Donnarumma}, {Tavecchio}, {Liodakis}, {Kiehlmann}, {Agudo}, {Jorstad}, {Muleri}, {Marscher}, {Puccetti}, {Middei}, {Perri}, {Pacciani}, {Negro}, {Romani}, {Di Marco}, {Blinov}, {Bourbah}, {Kontopodis}, {Mandarakas}, {Romanopoulos}, {Skalidis}, {Vervelaki}, {Casadio}, {Escudero}, {Myserlis}, {Gurwell}, {Rao}, {Keating}, {Kouch}, {Lindfors}, {Aceituno}, {Bernardos}, {Bonnoli}, {Casanova}, {Garc{\'\i}a-Comas}, {Ag{\'\i}s-Gonz{\'a}lez}, {Husillos}, {Marchini}, {Sota}, {Imazawa}, {Sasada}, {Fukazawa}, {Kawabata}, {Uemura}, {Mizuno}, {Nakaoka}, {Akitaya}, {Savchenko}, {Vasilyev}, {G{\'o}mez}, {Antonelli}, {Barnouin}, {Bonino}, {Cavazzuti}, {Costamante}, {Chen}, {Cibrario}, {De Rosa}, {Di Pierro}, {Errando}, {Kaaret}, {Karas}, {Krawczynski}, {Lisalda}, {Madejski}, {Malacaria}, {Marin}, {Marinucci}, {Massaro}, {Matt}, {Mitsuishi}, {O'Dell}, {Paggi}, {Peirson}, {Petrucci}, {Ramsey}, {Tennant}, {Wu}, {Bachetti}, {Baldini}, {Baumgartner},
  {Bellazzini}, {Bianchi}, {Bongiorno}, {Brez}, {Bucciantini}, {Capitanio}, {Castellano}, {Ciprini}, {Costa}, {Del Monte}, {Di Lalla}, {Doroshenko}, {Dov{\v{c}}iak}, {Enoto}, {Evangelista}, {Fabiani}, {Ferrazzoli}, {Garcia}, {Gunji}, {Hayashida}, {Heyl}, {Iwakiri}, {Kislat}, {Kitaguchi}, {Kolodziejczak}, {La Monaca}, {Latronico}, {Maldera}, {Manfreda}, {Ng}, {Omodei}, {Oppedisano}, {Papitto}, {Pavlov}, {Pesce-Rollins}, {Pilia}, {Possenti}, {Poutanen}, {Rankin}, {Ratheesh}, {Roberts}, {Sgr{\`o}}, {Slane}, {Soffitta}, {Spandre}, {Swartz}, {Tamagawa}, {Taverna}, {Tawara}, {Thomas}, {Tombesi}, {Trois}, {Tsygankov}, {Turolla}, {Vink}, {Weisskopf}, {Xie}, \& {Zane}}]{DiGesu2023}
{Di Gesu}, L., {Marshall}, H.~L., {Ehlert}, S.~R., {et~al.} 2023, Nature Astronomy, 7, 1245, \dodoi{10.1038/s41550-023-02032-7}

\bibitem[{{IceCube Collaboration} {et~al.}(2018){IceCube Collaboration}, {Aartsen}, {Ackermann}, {Adams}, {Aguilar}, {Ahlers}, {Ahrens}, {Samarai}, {Altmann}, {Andeen}, \& et~al.}]{IceCube2018}
{IceCube Collaboration}, {Aartsen}, M.~G., {Ackermann}, M., {et~al.} 2018, Science, 361, 147, \dodoi{10.1126/science.aat2890}

\bibitem[{{Itoh} {et~al.}(2016){Itoh}, {Nalewajko}, {Fukazawa}, {Uemura}, {Tanaka}, {Kawabata}, {Madejski}, {Schinzel}, {Kanda}, {Shiki}, {Akitaya}, {Kawabata}, {Moritani}, {Nakaoka}, {Ohsugi}, {Sasada}, {Takaki}, {Takata}, {Ui}, {Yamanaka}, \& {Yoshida}}]{Itoh2016}
{Itoh}, R., {Nalewajko}, K., {Fukazawa}, Y., {et~al.} 2016, \apj, 833, 77, \dodoi{10.3847/1538-4357/833/1/77}

\bibitem[{{Jorstad} {et~al.}(2017){Jorstad}, {Marscher}, {Morozova}, {Troitsky}, {Agudo}, {Casadio}, {Foord}, {G{\'o}mez}, {MacDonald}, {Molina}, {L{\"a}hteenm{\"a}ki}, {Tammi}, \& {Tornikoski}}]{Jorstad2017}
{Jorstad}, S.~G., {Marscher}, A.~P., {Morozova}, D.~A., {et~al.} 2017, \apj, 846, 98, \dodoi{10.3847/1538-4357/aa8407}

\bibitem[{{Keivani} {et~al.}(2018){Keivani}, {Murase}, {Petropoulou}, {Fox}, {Cenko}, {Chaty}, {Coleiro}, {DeLaunay}, {Dimitrakoudis}, {Evans}, {Kennea}, {Marshall}, {Mastichiadis}, {Osborne}, {Santander}, {Tohuvavohu}, \& {Turley}}]{Keivani2018}
{Keivani}, A., {Murase}, K., {Petropoulou}, M., {et~al.} 2018, \apj, 864, 84, \dodoi{10.3847/1538-4357/aad59a}

\bibitem[{{Krawczynski}(2012)}]{Krawczynski2012}
{Krawczynski}, H. 2012, \apj, 744, 30, \dodoi{10.1088/0004-637X/744/1/30}

\bibitem[{{Liodakis} {et~al.}(2022{\natexlab{a}}){Liodakis}, {Blinov}, {Potter}, \& {Rieger}}]{Liodakis2022-II}
{Liodakis}, I., {Blinov}, D., {Potter}, S.~B., \& {Rieger}, F.~M. 2022{\natexlab{a}}, \mnras, 509, L21, \dodoi{10.1093/mnrasl/slab118}

\bibitem[{{Liodakis} {et~al.}(2017){Liodakis}, {Pavlidou}, {Hovatta}, {Max-Moerbeck}, {Pearson}, {Richards}, \& {Readhead}}]{Liodakis2017}
{Liodakis}, I., {Pavlidou}, V., {Hovatta}, T., {et~al.} 2017, \mnras, 467, 4565, \dodoi{10.1093/mnras/stx432}

\bibitem[{{Liodakis} {et~al.}(2019{\natexlab{a}}){Liodakis}, {Peirson}, \& {Romani}}]{Liodakis2019-II}
{Liodakis}, I., {Peirson}, A.~L., \& {Romani}, R.~W. 2019{\natexlab{a}}, \apj, 880, 29, \dodoi{10.3847/1538-4357/ab2719}

\bibitem[{{Liodakis} \& {Petropoulou}(2020)}]{Liodakis2020}
{Liodakis}, I., \& {Petropoulou}, M. 2020, \apjl, 893, L20, \dodoi{10.3847/2041-8213/ab830a}

\bibitem[{{Liodakis} {et~al.}(2018){Liodakis}, {Romani}, {Filippenko}, {Kiehlmann}, {Max-Moerbeck}, {Readhead}, \& {Zheng}}]{Liodakis2018}
{Liodakis}, I., {Romani}, R.~W., {Filippenko}, A.~V., {et~al.} 2018, \mnras, 480, 5517, \dodoi{10.1093/mnras/sty2264}

\bibitem[{{Liodakis} {et~al.}(2019{\natexlab{b}}){Liodakis}, {Romani}, {Filippenko}, {Kocevski}, \& {Zheng}}]{Liodakis2019}
{Liodakis}, I., {Romani}, R.~W., {Filippenko}, A.~V., {Kocevski}, D., \& {Zheng}, W. 2019{\natexlab{b}}, \apj, 880, 32, \dodoi{10.3847/1538-4357/ab26b7}

\bibitem[{{Liodakis} {et~al.}(2020){Liodakis}, {Blinov}, {Jorstad}, {Arkharov}, {Di Paola}, {Efimova}, {Grishina}, {Kiehlmann}, {Kopatskaya}, {Larionov}, {Larionova}, {Larionova}, {Marscher}, {Morozova}, {Nikiforova}, {Pavlidou}, {Traianou}, {Troitskaya}, {Troitsky}, {Uemura}, \& {Weaver}}]{Liodakis2020-II}
{Liodakis}, I., {Blinov}, D., {Jorstad}, S.~G., {et~al.} 2020, \apj, 902, 61, \dodoi{10.3847/1538-4357/abb1b8}

\bibitem[{{Liodakis} {et~al.}(2022{\natexlab{b}}){Liodakis}, {Marscher}, {Agudo}, {Berdyugin}, {Bernardos}, {Bonnoli}, {Borman}, {Casadio}, {Casanova}, {Cavazzuti}, {Rodriguez Cavero}, {Di Gesu}, {Di Lalla}, {Donnarumma}, {Ehlert}, {Errando}, {Escudero}, {Garc{\'\i}a-Comas}, {Ag{\'\i}s-Gonz{\'a}lez}, {Husillos}, {Jormanainen}, {Jorstad}, {Kagitani}, {Kopatskaya}, {Kravtsov}, {Krawczynski}, {Lindfors}, {Larionova}, {Madejski}, {Marin}, {Marchini}, {Marshall}, {Morozova}, {Massaro}, {Masiero}, {Mawet}, {Middei}, {Millar-Blanchaer}, {Myserlis}, {Negro}, {Nilsson}, {O'Dell}, {Omodei}, {Pacciani}, {Paggi}, {Panopoulou}, {Peirson}, {Perri}, {Petrucci}, {Poutanen}, {Puccetti}, {Romani}, {Sakanoi}, {Savchenko}, {Sota}, {Tavecchio}, {Tinyanont}, {Vasilyev}, {Weaver}, {Zhovtan}, {Antonelli}, {Bachetti}, {Baldini}, {Baumgartner}, {Bellazzini}, {Bianchi}, {Bongiorno}, {Bonino}, {Brez}, {Bucciantini}, {Capitanio}, {Castellano}, {Ciprini}, {Costa}, {De Rosa}, {Del Monte}, {Di Marco}, {Doroshenko}, {Dov{\v{c}}iak}, {Enoto},
  {Evangelista}, {Fabiani}, {Ferrazzoli}, {Garcia}, {Gunji}, {Hayashida}, {Heyl}, {Iwakiri}, {Karas}, {Kitaguchi}, {Kolodziejczak}, {La Monaca}, {Latronico}, {Maldera}, {Manfreda}, {Marinucci}, {Matt}, {Mitsuishi}, {Mizuno}, {Muleri}, {Ng}, {Oppedisano}, {Papitto}, {Pavlov}, {Pesce-Rollins}, {Pilia}, {Possenti}, {Ramsey}, {Rankin}, {Ratheesh}, {Sgr{\'o}}, {Slane}, {Soffitta}, {Spandre}, {Tamagawa}, {Taverna}, {Tawara}, {Tennant}, {Thomas}, {Tombesi}, {Trois}, {Tsygankov}, {Turolla}, {Vink}, {Weisskopf}, {Wu}, {Xie}, \& {Zane}}]{Liodakis2022}
{Liodakis}, I., {Marscher}, A.~P., {Agudo}, I., {et~al.} 2022{\natexlab{b}}, \nat, 611, 677, \dodoi{10.1038/s41586-022-05338-0}

\bibitem[{{Lyutikov} {et~al.}(2005){Lyutikov}, {Pariev}, \& {Gabuzda}}]{Lyutikov2005}
{Lyutikov}, M., {Pariev}, V.~I., \& {Gabuzda}, D.~C. 2005, \mnras, 360, 869, \dodoi{10.1111/j.1365-2966.2005.08954.x}

\bibitem[{{Mannheim} \& {Biermann}(1992)}]{Mannheim1992}
{Mannheim}, K., \& {Biermann}, P.~L. 1992, \aap, 253, L21

\bibitem[{{Maraschi} {et~al.}(1992){Maraschi}, {Ghisellini}, \& {Celotti}}]{Maraschi1992}
{Maraschi}, L., {Ghisellini}, G., \& {Celotti}, A. 1992, \apjl, 397, L5, \dodoi{10.1086/186531}

\bibitem[{{Marscher}(2014)}]{Marscher2014}
{Marscher}, A.~P. 2014, \apj, 780, 87, \dodoi{10.1088/0004-637X/780/1/87}

\bibitem[{{Marscher} \& {Gear}(1985)}]{Marscher1985}
{Marscher}, A.~P., \& {Gear}, W.~K. 1985, \apj, 298, 114, \dodoi{10.1086/163592}

\bibitem[{{Marscher} {et~al.}(2010){Marscher}, {Jorstad}, {Larionov}, {Aller}, {Aller}, {L{\"a}hteenm{\"a}ki}, {Agudo}, {Smith}, {Gurwell}, {Hagen-Thorn}, {Konstantinova}, {Larionova}, {Larionova}, {Melnichuk}, {Blinov}, {Kopatskaya}, {Troitsky}, {Tornikoski}, {Hovatta}, {Schmidt}, {D'Arcangelo}, {Bhattarai}, {Taylor}, {Olmstead}, {Manne-Nicholas}, {Roca-Sogorb}, {G{\'o}mez}, {McHardy}, {Kurtanidze}, {Nikolashvili}, {Kimeridze}, \& {Sigua}}]{Marscher2010}
{Marscher}, A.~P., {Jorstad}, S.~G., {Larionov}, V.~M., {et~al.} 2010, \apjl, 710, L126, \dodoi{10.1088/2041-8205/710/2/L126}

\bibitem[{{Marshall} {et~al.}(2023){Marshall}, {Liodakis}, {Marscher}, {Di Lalla}, {Jorstad}, {Kim}, {Middei}, {Negro}, {Omodei}, {Peirson}, {Perri}, {Puccetti}, {Agudo}, {Bonnoli}, {Berdyugin}, {Cavazzuti}, {Rodriguez Cavero}, {Donnarumma}, {Di Gesu}, {Jormanainen}, {Krawczynski}, {Lindfors}, {Marin}, {Massaro}, {Pacciani}, {Poutanen}, {Tavecchio}, {Kouch}, {Aceituno}, {Bernardos}, {Bonnoli}, {Casanova}, {Garcia-Comas}, {Agis-Gonzalez}, {Husillos}, {Marchini}, {Sota}, {Blinov}, {Bourbah}, {Kielhmann}, {Kontopodis}, {Mandarakas}, {Romanopoulos}, {Skalidis}, {Vervelaki}, {Borman}, {Kopatskaya}, {Larionova}, {Morozova}, {Savchenko}, {Vasilyev}, {Zhovtan}, {Casadio}, {Escudero}, {Kramer}, {Myserlis}, {Trainou}, {Imazawa}, {Sasada}, {Fukazawa}, {Kawabata}, {Uemura}, {Mizuno}, {Nakaoka}, {Akitaya}, {Masiero}, {Mawet}, {Millar-Blanchaer}, {Panopoulou}, {Tinyanont}, {Berdyugin}, {Kagitani}, {Kravtsov}, {Sakanoi}, {Antonelli}, {Bachetti}, {Baldini}, {Baumgartner}, {Bellazzini}, {Bianchi}, {Bongiorno}, {Bonino},
  {Brez}, {Bucciantini}, {Capitanio}, {Castellano}, {Cavazzuti}, {Chen}, {Ciprini}, {Costa}, {De Rosa}, {Del Monte}, {Di Gesu}, {Di Marco}, {Donnarumma}, {Doroshenko}, {Dovvciak}, {Ehlert}, {Enoto}, {Evangelista}, {Fabiani}, {Ferrazzoli}, {Garcia}, {Gunji}, {Hayashida}, {Heyl}, {Iwakiri}, {Kaaret}, {Karas}, {Kislat}, {Kitaguchi}, {Kolodziejczak}, {Krawczynski}, {La Monaca}, {Latronico}, {Maldera}, {Manfreda}, {Marin}, {Marinucci}, {Matt}, {Mitsuishi}, {Mizuno}, {Muleri}, {Ng}, {ODell}, {Oppedisano}, {Papitto}, {Pavlov}, {Pesce-Rollins}, {Petrucci}, {Pilia}, {Possenti}, {Poutanen}, {Puccetti}, {Ramsey}, {Rankin}, {Ratheesh}, {Roberts}, {Romani}, {Sgro}, {Slane}, {Soffitta}, {Spandre}, {Swartz}, {Tamagawa}, {Taverna}, {Tawara}, {Tennant}, {Thomas}, {Tombesi}, {Trois}, {Tsygankov}, {Turolla}, {Vink}, {Weisskopf}, {Wu}, {Xie}, \& {Zane}}]{Marshall2023}
{Marshall}, H.~L., {Liodakis}, I., {Marscher}, A.~P., {et~al.} 2023, arXiv e-prints, arXiv:2310.11510, \dodoi{10.48550/arXiv.2310.11510}

\bibitem[{{Middei} {et~al.}(2023){Middei}, {Perri}, {Puccetti}, {Liodakis}, {Di Gesu}, {Marscher}, {Rodriguez Cavero}, {Tavecchio}, {Donnarumma}, {Laurenti}, {Jorstad}, {Agudo}, {Marshall}, {Pacciani}, {Kim}, {Aceituno}, {Bonnoli}, {Casanova}, {Ag{\'\i}s-Gonz{\'a}lez}, {Sota}, {Casadio}, {Escudero}, {Myserlis}, {Sievers}, {Kouch}, {Lindfors}, {Gurwell}, {Keating}, {Rao}, {Kang}, {Lee}, {Kim}, {Cheong}, {Jeong}, {Angelakis}, {Kraus}, {Antonelli}, {Bachetti}, {Baldini}, {Baumgartner}, {Bellazzini}, {Bianchi}, {Bongiorno}, {Bonino}, {Brez}, {Bucciantini}, {Capitanio}, {Castellano}, {Cavazzuti}, {Chen}, {Ciprini}, {Costa}, {De Rosa}, {Del Monte}, {Di Lalla}, {Di Marco}, {Doroshenko}, {Dov{\v{c}}iak}, {Ehlert}, {Enoto}, {Evangelista}, {Fabiani}, {Ferrazzoli}, {Garc{\'\i}a}, {Gunji}, {Hayashida}, {Heyl}, {Iwakiri}, {Kaaret}, {Karas}, {Kislat}, {Kitaguchi}, {Kolodziejczak}, {Krawczynski}, {La Monaca}, {Latronico}, {Maldera}, {Manfreda}, {Marin}, {Marinucci}, {Massaro}, {Matt}, {Mitsuishi}, {Mizuno}, {Muleri},
  {Negro}, {Ng}, {O'Dell}, {Omodei}, {Oppedisano}, {Papitto}, {Pavlov}, {Peirson}, {Pesce-Rollins}, {Petrucci}, {Pilia}, {Possenti}, {Poutanen}, {Ramsey}, {Rankin}, {Ratheesh}, {Roberts}, {Romani}, {Sgr{\`o}}, {Slane}, {Soffitta}, {Spandre}, {Swartz}, {Tamagawa}, {Taverna}, {Tawara}, {Tennant}, {Thomas}, {Tombesi}, {Trois}, {Tsygankov}, {Turolla}, {Vink}, {Weisskopf}, {Wu}, {Xie}, \& {Zane}}]{Middei2023}
{Middei}, R., {Perri}, M., {Puccetti}, S., {et~al.} 2023, \apjl, 953, L28, \dodoi{10.3847/2041-8213/acec3e}

\bibitem[{{M{\"u}cke} \& {Protheroe}(2001)}]{Mucke2001}
{M{\"u}cke}, A., \& {Protheroe}, R.~J. 2001, Astroparticle Physics, 15, 121, \dodoi{10.1016/S0927-6505(00)00141-9}

\bibitem[{{Paliya} {et~al.}(2018){Paliya}, {Zhang}, {B{\"o}ttcher}, {Ajello}, {Dom{\'\i}nguez}, {Joshi}, {Hartmann}, \& {Stalin}}]{Paliya2018}
{Paliya}, V.~S., {Zhang}, H., {B{\"o}ttcher}, M., {et~al.} 2018, \apj, 863, 98, \dodoi{10.3847/1538-4357/aad1f0}

\bibitem[{{Peirson} {et~al.}(2022){Peirson}, {Liodakis}, \& {Romani}}]{Peirson2022}
{Peirson}, A.~L., {Liodakis}, I., \& {Romani}, R.~W. 2022, \apj, 931, 59, \dodoi{10.3847/1538-4357/ac6a54}

\bibitem[{{Peirson} \& {Romani}(2019)}]{Peirson2019}
{Peirson}, A.~L., \& {Romani}, R.~W. 2019, \apj, 885, 76, \dodoi{10.3847/1538-4357/ab46b1}

\bibitem[{{Petropoulou} {et~al.}(2020){Petropoulou}, {Murase}, {Santander}, {Buson}, {Tohuvavohu}, {Kawamuro}, {Vasilopoulos}, {Negoro}, {Ueda}, {Siegel}, {Keivani}, {Kawai}, {Mastichiadis}, \& {Dimitrakoudis}}]{Petropoulou2020}
{Petropoulou}, M., {Murase}, K., {Santander}, M., {et~al.} 2020, \apj, 891, 115, \dodoi{10.3847/1538-4357/ab76d0}

\bibitem[{{Rani} {et~al.}(2013){Rani}, {Krichbaum}, {Fuhrmann}, {B{\"o}ttcher}, {Lott}, {Aller}, {Aller}, {Angelakis}, {Bach}, {Bastieri}, {Falcone}, {Fukazawa}, {Gabanyi}, {Gupta}, {Gurwell}, {Itoh}, {Kawabata}, {Krips}, {L{\"a}hteenm{\"a}ki}, {Liu}, {Marchili}, {Max-Moerbeck}, {Nestoras}, {Nieppola}, {Quintana-Lacaci}, {Readhead}, {Richards}, {Sasada}, {Sievers}, {Sokolovsky}, {Stroh}, {Tammi}, {Tornikoski}, {Uemura}, {Ungerechts}, {Urano}, \& {Zensus}}]{Rani2013}
{Rani}, B., {Krichbaum}, T.~P., {Fuhrmann}, L., {et~al.} 2013, \aap, 552, A11, \dodoi{10.1051/0004-6361/201321058}

\bibitem[{{Scarpa} \& {Falomo}(1997)}]{Scarpa1997}
{Scarpa}, R., \& {Falomo}, R. 1997, \aap, 325, 109

\bibitem[{{Schutte} {et~al.}(2022){Schutte}, {Britto}, {B{\"o}ttcher}, {van Soelen}, {Marais}, {Kaur}, {Falcone}, {Buckley}, {Rajoelimanana}, \& {Cooper}}]{Schutte2022}
{Schutte}, H.~M., {Britto}, R.~J., {B{\"o}ttcher}, M., {et~al.} 2022, \apj, 925, 139, \dodoi{10.3847/1538-4357/ac3cb5}

\bibitem[{{Sikora} {et~al.}(1994){Sikora}, {Begelman}, \& {Rees}}]{Sikora1994}
{Sikora}, M., {Begelman}, M.~C., \& {Rees}, M.~J. 1994, \apj, 421, 153, \dodoi{10.1086/173633}

\bibitem[{{Smith} {et~al.}(2007){Smith}, {Williams}, {Schmidt}, {Diamond-Stanic}, \& {Means}}]{Smith2007}
{Smith}, P.~S., {Williams}, G.~G., {Schmidt}, G.~D., {Diamond-Stanic}, A.~M., \& {Means}, D.~L. 2007, \apj, 663, 118, \dodoi{10.1086/517992}

\bibitem[{{Zdziarski} \& {B\"ottcher}(2015)}]{Zdziarski2015}
{Zdziarski}, A.~A., \& {B\"ottcher}, M. 2015, \mnras, 450, L21, \dodoi{10.1093/mnrasl/slv039}

\bibitem[{{Zhang}(2019)}]{ZHC2019b}
{Zhang}, H. 2019, Galaxies, 7, 85, \dodoi{10.3390/galaxies7040085}

\bibitem[{{Zhang} \& {B{\"o}ttcher}(2013)}]{ZHC2013}
{Zhang}, H., \& {B{\"o}ttcher}, M. 2013, \apj, 774, 18, \dodoi{10.1088/0004-637X/774/1/18}

\bibitem[{{Zhang} {et~al.}(2015){Zhang}, {Chen}, {B{\"o}ttcher}, {Guo}, \& {Li}}]{ZHC2015}
{Zhang}, H., {Chen}, X., {B{\"o}ttcher}, M., {Guo}, F., \& {Li}, H. 2015, \apj, 804, 58, \dodoi{10.1088/0004-637X/804/1/58}

\bibitem[{{Zhang} {et~al.}(2016){Zhang}, {Diltz}, \& {B{\"o}ttcher}}]{ZHC2016b}
{Zhang}, H., {Diltz}, C., \& {B{\"o}ttcher}, M. 2016, \apj, 829, 69, \dodoi{10.3847/0004-637X/829/2/69}

\bibitem[{{Zhang} {et~al.}(2019{\natexlab{a}}){Zhang}, {Fang}, {Li}, {Giannios}, {B{\"o}ttcher}, \& {Buson}}]{ZHC2019}
{Zhang}, H., {Fang}, K., {Li}, H., {et~al.} 2019{\natexlab{a}}, \apj, 876, 109, \dodoi{10.3847/1538-4357/ab158d}

\bibitem[{{Zhang} {et~al.}(2020){Zhang}, {Li}, {Giannios}, {Guo}, {Liu}, \& {Dong}}]{ZHC2020}
{Zhang}, H., {Li}, X., {Giannios}, D., {et~al.} 2020, \apj, 901, 149, \dodoi{10.3847/1538-4357/abb1b0}

\bibitem[{{Zhang} {et~al.}(2018){Zhang}, {Li}, {Guo}, \& {Giannios}}]{ZHC2018}
{Zhang}, H., {Li}, X., {Guo}, F., \& {Giannios}, D. 2018, \apjl, 862, L25, \dodoi{10.3847/2041-8213/aad54f}

\bibitem[{{Zhang} {et~al.}(2023){Zhang}, {Marscher}, {Guo}, {Giannios}, {Li}, \& {Negro}}]{ZHC2023}
{Zhang}, H., {Marscher}, A.~P., {Guo}, F., {et~al.} 2023, \apj, 949, 71, \dodoi{10.3847/1538-4357/acc657}

\bibitem[{{Zhang} {et~al.}(2019{\natexlab{b}}){Zhang}, {Santangelo}, {Feroci}, {Xu}, {Lu}, {Chen}, {Feng}, {Zhang}, {Brandt}, {Hernanz}, {Baldini}, {Bozzo}, {Campana}, {De Rosa}, {Dong}, {Evangelista}, {Karas}, {Meidinger}, {Meuris}, {Nandra}, {Pan}, {Pareschi}, {Orleanski}, {Huang}, {Schanne}, {Sironi}, {Spiga}, {Svoboda}, {Tagliaferri}, {Tenzer}, {Vacchi}, {Zane}, {Walton}, {Wang}, {Winter}, {Wu}, {in't Zand}, {Ahangarianabhari}, {Ambrosi}, {Ambrosino}, {Barbera}, {Basso}, {Bayer}, {Bellazzini}, {Bellutti}, {Bertucci}, {Bertuccio}, {Borghi}, {Cao}, {Cadoux}, {Campana}, {Ceraudo}, {Chen}, {Chen}, {Chevenez}, {Civitani}, {Cui}, {Cui}, {Dauser}, {Del Monte}, {Di Cosimo}, {Diebold}, {Doroshenko}, {Dovciak}, {Du}, {Ducci}, {Fan}, {Favre}, {Fuschino}, {G{\'a}lvez}, {Gao}, {Ge}, {Gevin}, {Grassi}, {Gu}, {Gu}, {Han}, {Hong}, {Hu}, {Ji}, {Jia}, {Jiang}, {Kennedy}, {Kreykenbohm}, {Kuvvetli}, {Labanti}, {Latronico}, {Li}, {Li}, {Li}, {Li}, {Li}, {Limousin}, {Liu}, {Liu}, {Lu}, {Luo}, {Macera}, {Malcovati},
  {Martindale}, {Michalska}, {Meng}, {Minuti}, {Morbidini}, {Muleri}, {Paltani}, {Perinati}, {Picciotto}, {Piemonte}, {Qu}, {Rachevski}, {Rashevskaya}, {Rodriguez}, {Schanz}, {Shen}, {Sheng}, {Song}, {Song}, {Sgro}, {Sun}, {Tan}, {Uttley}, {Wang}, {Wang}, {Wang}, {Wang}, {Wang}, {Wang}, {Watts}, {Wen}, {Wilms}, {Xiong}, {Yang}, {Yang}, {Yang}, {Yu}, {Zhang}, {Zampa}, {Zampa}, {Zdziarski}, {Zhang}, {Zhang}, {Zhang}, {Zhang}, {Zhang}, {Zhang}, {Zhang}, {Zhang}, {Zhao}, {Zheng}, {Zhou}, {Zorzi}, \& {Zwart}}]{extp2019}
{Zhang}, S., {Santangelo}, A., {Feroci}, M., {et~al.} 2019{\natexlab{b}}, Science China Physics, Mechanics, and Astronomy, 62, 29502, \dodoi{10.1007/s11433-018-9309-2}

\end{thebibliography}
\bibliographystyle{aasjournal}



\end{document}